\documentclass[12pt,thmsa]{article}
\usepackage{sw20lart}

\input{tcilatex}

\begin{document}

\author{Sascha Vongehr \\
%EndAName
Department of Physics and Astronomy\\
University of Southern California\\
Los Angeles, CA 90089-0484\\
USA}
\title{Nature: "I have Two Times"}
\date{7.7.1999}
\maketitle

\begin{abstract}
This is a slightly extended version of a seminar given the 8th of June at
the TASI 99 at Colorado University in Boulder. The motivations behind two
time theory are explained and the theory is introduced via one of the
theory's easier gauges of a particle on a black hole background. Important
results that should be interesting as well in the light of the recent AdS
mania will be summarized.
\end{abstract}

\section{Motivations for Two Times}

\subsection{"I have Two Times"}

Here is one common misconception about two time theory: They saw many
spatial dimensions being used so they tried many time dimensions. After
realizing there would be new negative norm states (ghosts) they looked
desperately for a gauge freedom to fix it.

Seen this way, the theory seems artificial, contrived and just not asked for.

This is NOT the way two time theory has been found. I personally like to
introduce it in the following way (not quite historically stringent either)
- stressing that nature actually tells us that two times are needed:

The huge success of today's fundamental theories like the standard model and
general relativity is due to localization of global gauge freedoms. Say you
start out discovering that the world shows Poincare symmetry. Lifting this
symmetry to a local one (motto: why should a region here be too rigidly
related to a region over there) gives back general relativity. Once we
recognize that a global symmetry becomes much more fundamental when
localized, nature tells us in this way that it has gravity, photons, and so
on. There is one global symmetry that has never been localized. It is the $%
Sp(2)$ symmetry in phase space. The symmetry acts on the doublet 
\begin{equation}
\left( 
\begin{array}{c}
\Phi _{1} \\ 
\Phi _{2}
\end{array}
\right) =\left( 
\begin{array}{c}
X \\ 
P
\end{array}
\right)
\end{equation}
which lets us rewrite the quantization $\left[ X,P\right] =i\hbar $ as

\begin{equation}
\Phi _{i}g^{ij}\Phi _{j}=i\hbar 
\end{equation}
where $(g_{ij})=\left( 
\begin{array}{cc}
0 & -1 \\ 
1 & 0
\end{array}
\right) $. Just like nature tells us ''I have gravity'' when localizing the
Poincare symmetry, it says ''I have two times'' when localizing this
symmetry. Let us see how so. Write down the easiest Lagrangian that would
give us an energy $\dot{X}\cdot P$ back: 
\begin{equation}
\pounds =\dfrac{1}{2}\Phi ^{\dagger },_{\tau }g\Phi =\dfrac{1}{2}\Phi
_{i},_{\tau }g^{ij}\Phi _{j}
\end{equation}
where $(g_{ij})=\left( 
\begin{array}{cc}
0 & -1 \\ 
1 & 0
\end{array}
\right) $if $\Phi =\left( 
\begin{array}{c}
X \\ 
P
\end{array}
\right) $, or for example use the $Osp(1/2)$ group metric $(g_{ij})=\left( 
\begin{array}{ccc}
i & 0 & 0 \\ 
0 & 0 & -1 \\ 
0 & 1 & 0
\end{array}
\right) $ if $\Phi =\left( 
\begin{array}{c}
\Psi  \\ 
X \\ 
P
\end{array}
\right) $, that is, if we want to include spin degrees of freedom, or use $%
Sp(2N)$ for $N$ particles and so on. The localization is a standard method.
We introduce the covariant derivative $;_{\tau }=(,_{\tau }-gA)$ and get 
\begin{equation}
\pounds =\dfrac{1}{2}\Phi ^{\dagger };_{\tau }g\Phi =\dfrac{1}{2}\Phi
^{\dagger },_{\tau }g\Phi -\dfrac{1}{2}\Phi ^{\dagger }A\Phi 
\end{equation}
The Euler Lagrange equations give the equations of motion as: 
\begin{eqnarray}
\Phi ,_{\tau } &=&gA\Phi \Rightarrow \pounds ^{\dagger }=\pounds  \\
\Phi _{i}\Phi _{j} &=&0\Rightarrow X\cdot X=0=X\cdot P=0=P\cdot P
\label{eom}
\end{eqnarray}
Thus it follows that $X$ and $P$\ are two light like vectors that are not
parallel. Therefore they have to have two time like dimensions and they must
have $d+2$ entries. We write $X^{M}$ where $M\in \{0,0^{\prime },1,...\}$
and $X^{M}$ is now a $SO(d,2)$ vector. Thus we see via the localization that
nature insists on two times because this is the only way not to get a
trivial theory after gauge fixing all three gauge choices that $Sp(2)\simeq
Sl(2,R)\simeq SO(2,1)$ offers. The three gauge choices (five for $Osp(1/2)$)
fix a gauge surface inside the $SO(d,2)$ symmetric starting space such that
we are left with $SO(d-1,1)$. The third gauge choice fixes the
parametrization of the world line $\tau _{\left( t\right) }$. Note that we
have to have one more time and one more spatial direction - nature is
telling us about two times only, not about many times. We write $X^{M}$
where $M\in \{0,0^{\prime },1,1^{\prime },2,3,...,d-2\}$ . The ''accident'' $%
Sp(2)\simeq SO(2,1)$ means that we can interpret the above as conformal
gravity on the world line. For the $Osp(1/2$) case it is conformal
supergravity on the world line.

\subsection{More Motivations for Two Times}

Looking at M-theory dualities one immediately bumps into two problems:

Firstly, in order to put all the dualities into one group, we need a bigger
group than allowed by a $(d-1,1)$ signature of the space time. Looking at
supergravity is enough to understand this. People tried to find the one
group that could give us SUGRA, but $Osp(1/32)$ is too small and $Osp(1/64)$
would lead to particles with spin higher than $2$ if realized in the usual
signature. In order to embed all we know about M-theory and its dualities,
we need $Osp(1/64)$ and therefore we need two times if we want to avoid
particles with spin larger than $2$ . This means we need $13$ dimensions to
start with, i.e. $SO(11,2)$ to be gauged down to a $SO(10,1)$ theory.

Secondly, M-theory provides dualities between theories with differing
topologies. This is what two time theory is very strong in. Often different
theories are $Sp(2)$ -gauge duals with different topologies of the gauge
surfaces. Examples are the massless particle having a world line that goes
through the origin of the $X^{0}\times X^{0^{\prime }}$ -plane and the
simple harmonic oscillator having a world line that is a circle in this
plane.

\section{Remarks to be Noted}

Many people used spaces with two time like dimensions in one form or
another. There is a neat way to get AdS spaces from embeddings in two times.
These are geometrical tricks. The two time theory described here is very
different in the following ways:

1) It is a dynamical theory. We start out with an Lagrangian and can
generalize to the string case which means it is an interacting theory.

2) It gives well known ghostfree physical systems.

3) It is a consistent and unitary quantum theory. The $SO(d,2)$ covariant
quantization gives the Casimir $C_{2\left[ SO(d,2)\right] }=4C_{2\left[ Sp(2)%
\right] }-\dfrac{1}{4}(d^{2}-4)=0-\dfrac{1}{4}\left( d^{2}-4\right) $ and
the canonical or field theoretical quantization after gauge fixing gives $%
C_{2}=-\dfrac{1}{4}\left( d^{2}-4\right) $ also \cite{ib}.

4) We easily can include spin \cite{ibcd}\ and supersymmetry \cite{bdm}.

\section{Example}

It is high time to give an easy example. The example given will be published
shortly in a wider context \cite{von}. It is very easy since after gauge
fixing we are left with an only $1+1$-dimensional system. Write 
\begin{eqnarray}
M &=&\quad \left( +^{\prime }\quad -^{\prime }\quad 0\quad 1\right) \\
X^{M} &=&\quad \left( 1,\quad X^{-^{\prime }},\quad t,\,\quad r_{\ast
}\right) N_{(t,r)} \\
P^{M} &=&\dfrac{1}{N}\left( 0,\quad P^{-^{\prime }},\quad p_{t},\quad
p_{\ast }\right)
\end{eqnarray}
The $1$ and the $0$ are gauge choices, $N$ is a conformal factor, the rest
are names given. We have to satisfy $X\cdot X=0=X\cdot P$ (\ref{eom} ) and
do this with 
\begin{eqnarray}
M &=&\left( +^{\prime }\quad \quad -^{\prime }\quad \quad 0\quad \quad
1\right) \\
X^{M} &=&\left( 1,\quad \dfrac{1}{2}\left( r_{\ast }^{2}-t^{2}\right) ,\quad
t,\,\quad r_{\ast }\right) N_{(t,r)} \\
P^{M} &=&\dfrac{1}{N}\left( 0,\quad r_{\ast }p_{\ast }-tp_{t},\quad
p_{t},\quad p_{\ast }\right)
\end{eqnarray}
The metric is given by $\eta ^{+^{\prime }-^{\prime }}=-1$ and the line
element is 
\begin{equation}
(ds)^{2}=(dX^{M})(dX_{M})=N^{2}[-(dt)^{2}+(dr_{\ast })^{2}]
\end{equation}
The well known black hole metric 
\begin{equation}
ds^{2}=-N^{2}(dt)^{2}+N^{-2}(dr)^{2}
\end{equation}
requires $dr_{\ast }=drN^{-2}$. Thus we need to integrate to get the right $%
r_{\ast }$ from the $N$ that we want. We might like to model a particle on a
Reissner-Nordstrom background ($N=\sqrt{1-\dfrac{r_{+}}{r}}\sqrt{1-\dfrac{%
r_{-}}{r}}$) and get $r_{\ast }=r+\dfrac{r_{+}^{2}\ln \left( r-r_{+}\right)
+r_{-}^{2}\ln \left( r-r_{-}\right) }{r_{+}-r_{-}}$.

In order to show the quantum mechanical treatment etc. it is convenient to
define $\sqrt{2}u=t+r_{\ast }$and $\sqrt{2}v=t-r_{\ast }$: 
\begin{eqnarray}
M &=&\left( +^{\prime }\quad -^{\prime }\quad +\quad -\right) \\
X^{M} &=&\left( 1,\quad -uv,\quad u,\,\quad v\right) N_{(u,v)}  \nonumber \\
P^{M} &=&\dfrac{1}{N}\left( 0,\quad up_{u}+vp_{v},\quad -p_{v},\quad
-p_{u}\right) \\
dX^{M} &=&\left( 0,\quad -d\left( uv\right) ,\quad du,\quad dv\right) N-%
\frac{X^{M}}{N}dN
\end{eqnarray}
The metric is given by $\eta ^{+^{\prime }-^{\prime }}=\eta ^{+-}=-1$ and
the line element is 
\begin{equation}
(ds)^{2}=(dX^{M})(dX_{M})=-2N^{2}dudv
\end{equation}
Inserting these forms in the original Sp$\left( 2,R\right) $ local and SO$%
\left( 2,2\right) \,$global invariant Lagrangian \cite{ib}\ gives 
\begin{eqnarray}
\pounds &=&\dot{X}\cdot P-\frac{1}{2}A^{22}P\cdot P-\frac{1}{2}A^{11}X\cdot
X-A^{12}X\cdot P \\
&=&\dot{u}p_{u}+\dot{v}p_{v}+\dfrac{A^{22}}{N^{2}}p_{u}p_{v} \\
&=&\frac{-N^{2}}{A^{22}}\dot{u}\,\dot{v}=\frac{1}{2A^{22}}G_{\mu \nu }\dot{x}%
^{\mu }\dot{x}^{\nu }
\end{eqnarray}
The metric is recognized in the line element or in the last line $G_{\mu \nu
}=\eta _{\mu \nu }N^{2}$, which is obtained by integrating out the momenta.

This form shows that the system has the larger symmetry SO$\left( 2,2\right) 
$ whose generators are the Lorentz generators in the 2+2 dimensional space $%
L^{MN}=X^{M}P^{N}-X^{N}P^{M}$. We quantize via $[u,p_{u}]=i=[v,p_{v}]$. In
the present gauge these take a form that is quantum ordered already: 
\begin{eqnarray}
L^{+^{\prime }-^{\prime }} &=&up_{u}+vp_{v},\quad L^{+^{\prime }+}=-p_{v} \\
L^{+^{\prime }-} &=&-p_{u},\quad L^{-^{\prime }+}=-u^{2}p_{u} \\
L^{-^{\prime }-} &=&-v^{2}p_{v},\quad L^{+-}=-up_{u}+vp_{v}
\end{eqnarray}
Under the SO$\left( 2,2\right) =SL\left( 2,R\right) _{L}\otimes SL\left(
2,R\right) _{R}$ the generators may be reclassified in the form 
\begin{eqnarray}
G_{2}^{L} &=&vp_{v},\quad G_{+}^{L}=G_{0}^{L}+G_{1}^{L}=-p_{v},\quad
G_{-}^{L}=G_{0}^{L}-G_{1}^{L}=-v^{2}p_{v}\,\,, \\
G_{2}^{R} &=&up_{u},\quad G_{+}^{R}=G_{0}^{R}+G_{1}^{R}=-p_{u},\quad
G_{-}^{R}=G_{0}^{R}-G_{1}^{R}=-u^{2}p_{u}\,\,.
\end{eqnarray}
where $G_{0}^{L,R}$ are the compact generators and $G_{1,2}^{L,R}$ are the
non-compact ones.

The Casimir is $\ C_{2\text{ \ }}[SO(d,2)]=0$ as it should be in two
dimensions since $C_{2}=$ $1-\dfrac{d^{2}}{4}$.

\section{Results}

Two time theory led to some astonishing results already. Often the results
are ''just'' a better understanding or interpretation of known methods or
facts. Among these results fall the understanding of the origin of the $%
SO(d,2)$ -symmetry that occurs in certain spectra as being there from the
very start in the two time description.

There is the more natural interpretation of the standard method \cite{djg}
in order to go from global AdS to the black hole solution. One has had to
identify and complexify because both are gauges in a space time with two
times \cite{von}.

And for example the always mysterious kappa-symmetry has bosonic partners
and becomes understandable \cite{bdm}. Let us devide the main results into
two categories:

\subsection{Dualities}

In the point particle two time theory it was shown that the massless, the
massive, the relativistic, the non-relativistic particle, the simple
harmonic oscillator, the particle in an arbitrary central potential, the
particle on an $AdS$-background, on an $AdS\times S$-background, on an BTZ
background, on an Robertson-Bertotti background and others are all $Sp(2)$%
-gauge dual theories; meaning they all derive from the same two time
theoretical starting point via different choices of the three gauge
parameters. For the generalization to the string case where only the
tensionless and the rigid string have been found yet \cite{bdmstrings}, this
could mean that the tensionless and tensionful and IIA and IIB etc. strings
can all be gotten in a similar way from one common unifying theory with two
times.

\subsection{Hidden Symmetries}

The space-time before gauging shows $SO(d,2)$ -symmetry. This symmetry is
still there in the gauged theory but it is hard to see since it is
non-linearly realized. This was known only for the hydrogen atom and the
superparticle \cite{js}. In \cite{bdm} it was shown how the action of the
superparticle that has hidden superconformal symmetries derives from the two
time starting point where these symmetries are manifest and still linear
before projection onto the gauge surface. Two time theory has shown that
these hidden symmetries are inside all of the spectra of all the gauge dual
theories. That means that all the mentioned systems have these hidden
symmetries. ''Hidden'' means that the symmetries are symmetries of the
action and not of the Hamiltonian. Symmetries hidden in the action can be as
vital as the Lorentz boost symmetry for a free particle: A Lagrangian like $%
\pounds =-m\sqrt{1-\stackrel{.}{r}^{2}}$ (or better the Hamiltonian $H=\sqrt{%
p^{2}+m^{2}}$) does not reveal it but surely one would question someone's
understanding of the system at hand if that somebody were not to realize
that this expression describes a relativistic particle.

In the gauged SUGRA $AdS_{n}\times S^{m}$ the bosonic subgroup is $%
SO(n-1,2)\times SO(m+1)$ which is smaller than the full $SO(n+m,2)$. As
observed in \cite{ibt} already: The $AdS\times S$ discussion may benefit
from this.

\section{Outlook}

There are very many things that one might try to put into a two time
description. A few that have not been attempted yet are:

- The tensionless string gauge being found there should be a string with
tension. This is similar to the massless point particle gauge that was a
precursor for the massive one. Then of course it needs a generalization to
p-branes.

- The introduction of background fields is expected to lead to an
electromagnetic theory that is manifestly dual between electrical and
magnetic phenomena and still $SO(d,2)$ covariant of course as well.

- Only two time theory can give a large enough supergroup to give M-theory,
thus two time theory should lead to M-theory but there is only a toy-model 
\cite{bdmM} yet.

\subsection{How to get further into the subject}

This subject has been developed over the past few years and all the recent
papers by the pioneer of two time physics Itzhak Bars are recommendet. As a
further introduction all those papers can be a little too much though,
therefore there will be a different and self contained approach from first
principles given in \cite{von} .


\begin{thebibliography}{99}
\bibitem{von}  S. Vongehr, ''Black Holes In Two Times'' hep-th/9907nnn

\bibitem{btz}  M. Banados, C. Teitelboim and J. Zanelly, Phys.Rev.Lett. 69
(1992) 1849-1851

\bibitem{ibcd}  I. Bars and C. Deliduman,''Gauge symmetry in phase space
with spin'' hep-th/9806085

\bibitem{djg}  S. Deser, R. Jackiw and G't Hooft, Ann.Phys. 152 (1984) 220

\bibitem{ib}  I. Bars, ''Two-Time Physics'' hep-th/9809034

\bibitem{ibt}  I. Bars, ''Hidden Symmetries, AdS\_D x S\symbol{94}n, and the
lifting of one-time-physics to two-time-physics'' hep-th/9810025

\bibitem{bdm}  I. Bars, C. Deliduman, D. Minic, ''Supersymmetric Two-Time
Physics'' hep-th/9812161

\bibitem{bdmstrings}  I. Bars, C. Deliduman, D. Minic,''Strings, Branes and
Two-Time Physics'' hep-th/9906223

\bibitem{js}  J.H. Schwarz, Nucl. Phys.B 185 (1981) 221

\bibitem{bdmM}  I. Bars, C. Deliduman, D. Minic,''Lifting M-Theory to
Two-Time Physics'' hep-th/9904063
\end{thebibliography}
\end{document}